\documentclass[sigconf]{acmart} 

\AtBeginDocument{%
  \providecommand\BibTeX{{%
    \normalfont B\kern-0.5em{\scshape i\kern-0.25em b}\kern-0.8em\TeX}}}

\usepackage{soul}
\usepackage{xcolor}
\usepackage{booktabs} 
\usepackage{hhline}
\usepackage{array}
\usepackage{afterpage}
\usepackage{placeins}
\usepackage{color}
\usepackage{graphicx}
\usepackage{lipsum}
\usepackage{subfig}
\usepackage{caption}
\usepackage{tabularx}
\usepackage[nolist,nohyperlinks,smaller]{acronym}
\usepackage{tikz}
\usepackage{balance}

\definecolor{LIGHTGRAY}{HTML}{DDDDDD}

\copyrightyear{2024}
\acmYear{2024}
\setcopyright{acmlicensed}\acmConference[CHI '24]{Proceedings of the CHI Conference on Human Factors in Computing Systems}{May 11--16, 2024}{Honolulu, HI, USA}
\acmBooktitle{Proceedings of the CHI Conference on Human Factors in Computing Systems (CHI '24), May 11--16, 2024, Honolulu, HI, USA}
\acmDOI{10.1145/3613904.3642744}
\acmISBN{979-8-4007-0330-0/24/05}

\begin{document}

\title{Jigsaw: Authoring Immersive Storytelling Experiences with Augmented Reality and Internet of Things}

\author{Lei Zhang}
\affiliation{%
 \institution{University of Michigan}
 \city{Ann Arbor}
 \state{MI}
 \country{USA}}
 \email{raynez@umich.edu}

\author{Daekun Kim}
\affiliation{%
 \institution{University of Waterloo}
 \city{Waterloo}
 \state{ON}
 \country{Canada}}
 \email{d332kim@uwaterloo.ca}

\author{Youjean Cho}
\affiliation{%
 \institution{University of Washington}
 \city{Seattle}
 \state{WA}
 \country{USA}}
 \email{youjean.design@gmail.com}

\author{Ava Robinson}
\affiliation{%
 \institution{Snap Inc.}
 \city{Santa Monica}
 \state{CA}
 \country{USA}}
 \email{avamrobinson@gmail.com}

\author{Yu Jiang Tham}
\affiliation{%
 \institution{Snap Inc.}
 \city{Santa Monica}
 \state{CA}
 \country{USA}}
 \email{yujiang@snap.com}

\author{Rajan Vaish}
\affiliation{%
 \institution{Snap Inc.}
 \city{Santa Monica}
 \state{CA}
 \country{USA}}
 \email{vaish.rajan@gmail.com}

 \author{Andrés Monroy-Hernández}
\affiliation{%
 \institution{Snap Inc. and Princeton University}
 \city{Princeton}
 \state{NJ}
 \country{USA}}
 \email{andresmh@cs.princeton.edu}
\renewcommand{\shortauthors}{Zhang, et al.}

\newcommand{\lei}[1]{\textcolor{black}{#1}}
\newcommand{\dk}[1]{\textcolor{black}{#1}}
\newcommand{\yc}[1]{\textcolor{black}{#1}}
\begin{abstract}

Augmented Reality (AR) presents new opportunities for immersive storytelling. However, this immersiveness faces two main hurdles. First, AR's immersive quality is often confined to visual elements, such as pixels on a screen. Second, crafting immersive narratives is complex and generally beyond the reach of amateurs due to the need for advanced technical skills. We introduce Jigsaw, a system that empowers beginners to both experience and craft immersive stories, blending virtual and physical elements. Jigsaw uniquely combines mobile AR with readily available Internet-of-things (IoT) devices. We conducted a qualitative study with 20 participants to assess Jigsaw's effectiveness in both consuming and creating immersive narratives. The results were promising: participants not only successfully created their own immersive stories but also found the playback of three such stories deeply engaging. However, sensory overload emerged as a significant challenge in these experiences. We discuss design trade-offs and considerations for future endeavors in immersive storytelling involving AR and IoT.

\end{abstract}


\begin{teaserfigure}
  \includegraphics[width=\textwidth]{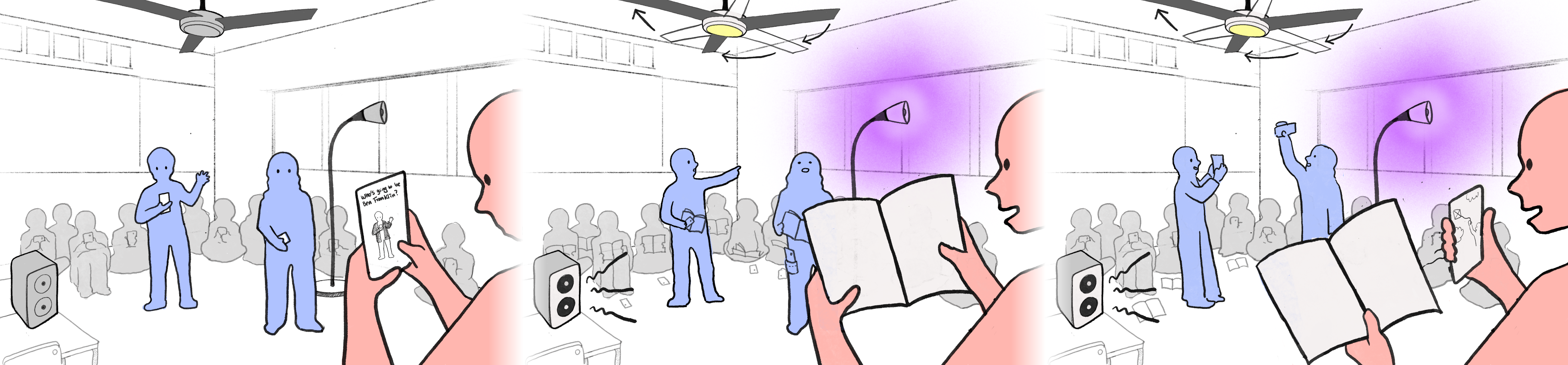}
  \caption{An example story of Benjamin Franklin's Kite Experiment from Jigsaw that combines AR and IoT devices into one immersive experience: (a) The participants can assign themselves a character by waving. (b) As the narrator reads the story, trigger words from the narration enact changes in the physical environment (shown with a smart light, smart fan, and smart speaker). (c) Virtual effects or objects are shown in the AR view, such as a kite, clouds, and sparks.}
  \Description{An example story of Benjamin Franklin's Kite Experiment from Jigsaw that combines AR and IoT devices into one immersive experience: (a) The participants can assign themselves a character by waving. (b) As the narrator reads the story, trigger words from the narration enact changes in the physical environment (shown with a smart light, smart fan, and smart speaker). (c) Virtual effects or objects are shown in the AR view, such as a kite, clouds, and sparks.}
  \label{fig:teaser}
  \vspace{1pc}
\end{teaserfigure}

\maketitle

\section{Introduction}

As technology develops, the media used for storytelling evolves too, from traditional methods, such as printed books and spoken tales, to digital formats like photos, videos, and animations. A key goal in these developments is to make stories more immersive and engaging. Augmented Reality (AR) is one such technology. AR opens new opportunities for immersive storytelling by showing 3D content in a spatial manner and allowing people to interact directly with it. However, AR's immersiveness is typically limited to pixels on a display. 

Beyond visuals, storytellers in theatres and performing venues have explored making experiences more immersive by adding elements that stimulate multiple senses, including sight, sound, touch, and smell. This approach appears in, for example, ``4D'' movies at immersive movie theaters and amusement parks. Also, plays like ``Sleep No More'' enhance engagement by allowing audiences to move around the performance space and connect with actors, making the experience more interactive. These immersive experiences, though captivating, often require specialized setups limited by cost, skills, and locations.

We introduce Jigsaw, a novel system that democratizes creating and enjoying immersive, multi-sensory stories. Jigsaw uses common mobile AR devices---like iPhones or Android phones--and household IoT devices--- such as smart lights and fans. This setup allows Jigsaw to display virtual 3D content and mimic environmental effects, like wind and light changes. Unlike previous systems, Jigsaw does not require specialized IoT devices or advanced programming skills ~\cite{storymakar}. Jigsaw lets people author stories with multiple scenes. Drawing inspiration from slide-based presentation software like PowerPoint, people can add AR elements, such as 3D models and animations, and control IoT devices, like dimming lights or adjusting fan speed. They can set triggers, such as a spoken word or touch, to move to the next scene. Additionally, multiple people can experience the story, each from their own unique perspective.

We evaluated Jigsaw with 20 participants in 10 groups, creating three specific stories. We found that Jisgsaw's stories were immersive, engaging, and memorable. The authoring mode was user-friendly and versatile for creating various stories. Our study also highlighted some trade-offs of this new storytelling modality, including the tension between engagement and sensory overload. 

This work presents three key contributions: 1) an AR + IoT tool that makes it easier to create immersive stories; 2) three immersive stories that showcase the possibilities of this approach; 3) findings from a study that highlight the advantages and difficulties of this innovative design approach.
\section{Background and Related Work}

\subsection{Immersive Storytelling}

\dk{Immersion is an important aspect in any storytelling to engage the audience and to convey the emotion of the performance. Therefore, many fields of performing arts have adopted novel forms of performance to enhance immersion. }Immersive theater is a prominent example ``which use installations and expansive environments, which have mobile audiences, and which invite audience participation'' \cite{white_2012}. It can involve great levels of intimacy \cite{goed_2004, goed_2007, goed_2010} or put more distinction between the audience and the actors  \cite{resnick_saxton_angove_lane_2010, resnick_saxton_angove_lane_2012}. The experience also involves multiple spatial awareness, touch, and smell \cite{white_2012}. Sleep No More, an immersive theater performance based on Shakespeare's Macbeth, enables the audience to be embedded within the performance, inviting them to walk around the stage and touch the props \cite{sleep_no_more}. While there are some examples of technology used in this type of performance, e.g., headphones and speakers in Rotozaza's Autoteatro \cite{hampton_mercuriali_2020}, it is rather limited, with many available technologies like projection mapping, AR, and robotics rarely being used if ever. We envision that technology has a great potential to enhance immersion and give a sense of magical events occurring in the world around us, as seen in literary genres like magical realism \cite{marquez_1981, allende_1982}.

Outside of theater, we see more frequent use of technology in storytelling. For instance, the Weather Channel frequently uses AR in their weather forecast programs \cite{feldman_2019}. \dk{Additionally, theme parks, such as Disneyland, use projection mapping, robots, and mechanically moving rides to foster a high sense of immersion in the storytelling of their rides \cite{startours, spacemountain}.} Projection mapping, in particular, is a popular method for AR to convert a room or a building into a magical space filled with eye-catching visualizations (e.g., \cite{sketchocean, digitalgraffiti, siccardi_2022}). 
\lei{Recent work has also proposed methods that combine motion capture and head-mounted AR displays to create immersive films~\cite{gholap2023past}.}
However, these storytelling experiences require a team of expert designers and engineers to configure, along with an expensive set of equipment. Therefore, they are difficult to access for most people and can only be experienced in special venues. In contrast, our work aims to democratize immersive storytelling to everyday users.

The emergence of AR-enabled mobile devices created a platform to tackle this issue. Wonderscope \cite{wonderscope} is a prime example that uses mobile AR to recreate immersive theaters for children in their own rooms. Using this, children can be immersed in the magical storytelling experience, presented as ready-made narratives. \dk{SceneAR achieves a similar immersive storytelling experience but adds the ability to remix the stories \cite{scenear2021}.} There are also examples of AR use for educational purposes, such as teaching chemistry \cite{elements4d}. While these systems \lei{provide} a starting point and an inspiration for our work, they are largely limited to augmenting the world with pixels on a screen. This fundamentally limits the level of immersion compared to previous examples of immersive theater and higher-fidelity experiences that give a sense of magic throughout the entire surroundings. Rather, we want to incorporate both virtual and physical augmentations (``pixels and atoms'') to foster a more complete immersion. Therefore, we examine the current landscape of ubiquitous computing and the possibility of using smart IoT devices as a latent infrastructure of immersive storytelling.

\subsection{Ubiquitous Computing and Physical Augmentation}
With the growing popularity of smart homes and IoT devices, ubiquitous computing is becoming a part of everyone's day-to-day life. Part of this effort aims to bring virtual augmentation to the world around us. \dk{For instance, RoomAlive \cite{roomalive} and Room2Room \cite{pejsa2016room} achieve room-scale projection mapping with commercially available devices such as consumer-grade projectors and Microsoft Kinect.} Moreover, Olwal and Dementyev \cite{olwal_dementyev_2022} showcase an ambient display hidden under common surfaces that can be seen in household \lei{furniture}. 
\dk{Many other works demonstrate the democratization of AR to the masses \cite{kim2023view}, including its applications in domains such as e-learning~\cite{choi2023prevailing}, e-commerce~\cite{al2023creative}, and social change~\cite{silva2022understanding}.}
As for use cases that focus on storytelling, Healey et al. \cite{healey_2021} demonstrate that mixed-reality systems can enable remote storytelling and that it helps promote engagement among children.
\lei{Li et al. proposed an intelligent approach that automatically adapt AR narratives to contextually compatible locations~\cite{li2023location}.}

\dk{However, as discussed above, we wish to go beyond pixel-based augmentation and leverage devices that physically augment the environment. Examples include smart lights, smart fans, speakers, smart thermostats, vacuum robots, and wearable devices like smart watches or smart glasses. }Additionally, we want to seamlessly integrate the virtual and physical augmentations to create a holistic narrative. There are previous works that leverage conversational agents on smart speakers \cite{cook2021, lin2019}, robots \cite{hubbard2021, kory2014, dang2023cargamear}, and tangible interfaces \cite{sylla2013} to foster effective storytelling with physical augmentation, but they largely focus on one or a handful of IoT device types, and often do not incorporate virtual augmentation. Lin et al. \cite{lin2019} showcase a more comprehensive, low-cost system that recreates 4D theater experience using home appliances like the ones mentioned before. \dk{However, the system nonetheless does not incorporate immersive virtual augmentations like mobile AR or AR glasses, and the system's authoring environment is not suited for novice users, which fails to meet our goal of enabling anyone to create immersive experiences at home. StoryMakAR \cite{storymakar} enables children to create an immersive storytelling experience using robots and mobile AR, which aims to achieve similar goals as our work but still does not incorporate other physical devices like fans, lights, thermostats, etc., to create a truly immersive experience that transforms the entire surroundings. We examine existing tools with similar goals to envision how to empower anyone to author this type of experience.}

\subsection{Authoring Tools}
\dk{Much of the efforts of the HCI community focus on how to make the creation of digital content easy, which helped transform what was considered the ``expert-only'' task of programming to something that anyone can do.} Scratch \cite{scratch} is a prime \lei{example} that uses block-based programming to empower children to author their own programs and games. Since then, many commercial applications have sprung up (e.g., graph-based programming in Lens Studio \cite{lens_studio}, Blender shader graph \cite{blender_shader_editor}, Unreal Blueprints \cite{unreal_blueprints}) with similar goals in mind. \dk{This effort to achieve easy authoring of programmable content also naturally came to focus on AR applications as they emerged. Lee et al. \cite{lee2004immersive} shows one of the first examples of this, where they enable immersive authoring of AR applications using tangible UI, but only limited to on-screen virtual contents without physical augmentations. FlowMatic \cite{flowmatic} is a more recent work that uses virtual reality (VR) and graph-based programming for the in-situ authoring of interactive scenes in an immersive VR environment.} SceneAR \cite{scenear2021} specifically targets viewing and remixing of immersive storytelling experiences and uses mobile AR for immersive authoring of these contents. While these works \lei{inspire} the design of our authoring tool, they are limited to authoring virtual augmentations, whereas we aim to encompass physical augmentations as well.
In contrast, there are authoring tools that target physical IoT devices. 
\lei{Ivy \cite{ivy} enables users to define behaviors of smart devices and their relational events via visual dataflow programming in VR.
Our system takes inspiration from their spatial editing UI, and contributes to authoring behaviors of both virtual AR content and physical IoT devices rather than being limited only to the latter.}
Other works demonstrate the use of conversational agents \cite{kimandko2022} and proxemic and gestural interactions \cite{progesar} for authoring an IoT environment. \dk{These works, however, also limit their target augmentation to only the physical.} StoryMakAR \cite{storymakar} lets users author immersive stories that involve both physical devices and virtual objects, which is much closer to what we aim to achieve. \dk{However, as discussed previously, the system does not handle a variety of smart devices to provide a fully immersive experience. It also requires a custom-built robotics toolkit and does not leverage the existing IoT infrastructure that could unlock easily accessible multi-sensory immersion.}

\dk{Jigsaw builds upon these prior works to enable anyone to create and experience immersive storytelling that combines both virtual and physical augmentation within a cohesive interactive system by leveraging commonly available IoT infrastructure and AR-enabled devices.}
\lei{Jigsaw also contributes to prior work by allowing users to define triggers such as keywords and touch interactions, as well as authoring multiple scenes that compose compelling immersive storytelling experiences.}
\section{Jigsaw System}

\begin{table*}
  \caption{Design considerations for each immersive story.}
  \label{tab:design}
  \begin{tabular}{llll}
    \toprule
    \textbf{Story} & \textbf{IoT Augmentation} & \textbf{User Role} & \textbf{Interaction}\\
    \midrule
    Benjamin Franklin & Background & Narrators and actors & Narration, touching, moving in the space\\
    The Wind and The Sun & Foreground & Audience & Tapping on the phone\\
    Goodnight Moon & Foreground & Narrators and audience & Narration\\
  \bottomrule
  \end{tabular}
\end{table*}

We designed Jigsaw to enable novices to both consume and create immersive stories using a mix of mobile AR and off-the-shelf IoT devices.
Our research process was divided into two steps: exploring \textit{what} immersive stories are, and investigating \textit{how} to create these immersive stories.
We executed our research in this order since we believe it is necessary to understand what the possible resulting experiences \lei{in this paradigm} are before building an authoring system for creating these experiences.

For the first step, we used two motivating questions to guide our design of immersive storytelling experiences: ``how to seamlessly integrate both virtual and physical augmentation to these immersive stories?'' and ``what are the levels of participation for users in interacting with each other and with the stories?''
We identified \lei{three} design considerations for immersive stories that involve AR and IoT devices.
Based on these design considerations, we then conducted a \lei{series} of brainstorming sessions to discuss potential stories via sketches and low-fidelity prototypes.
Lastly, we prototyped three representative stories (i.e., Benjamin Franklin, The Wind and the Sun, and Goodnight Moon) that cover different aspects of the design considerations.
The design considerations for each story are specified in Table \ref{tab:design}.
During the prototyping process, the research team met regularly to discuss design iterations and ran pilot tests frequently with end-users outside of our team using the prototypes.

For the second step, we designed and built an authoring system for creating these immersive stories.
We have three design goals in mind: \textit{(i)} to lower the barrier of entry for novices to create these stories, \textit{(ii)} to make the system generalizable for creating various stories including the three representative stories, and \textit{(iii)} to make the system consistent with existing interaction modalities in mobile AR applications.
We drew inspiration from presentation tools such as Google Slides and enabled novices to create a linear storyline by creating a series of scenes.
Within each scene, users can edit the triggers for the scene and the behaviors of AR content and IoT devices, \lei{inspired} by existing event-trigger models (e.g. as seen in~\cite{storymakar}).
Finally, users can play the story that they create in the same system.

We detail the design and implementation of the immersive stories and the authoring system in the following sections.

\subsection{Jigsaw's Immersive Stories}

\begin{figure*}[t!]
  \centering
  \subfloat[Benjamin Franklin.\label{fig:ben_franklin}]{\includegraphics[clip,width=0.33\linewidth]{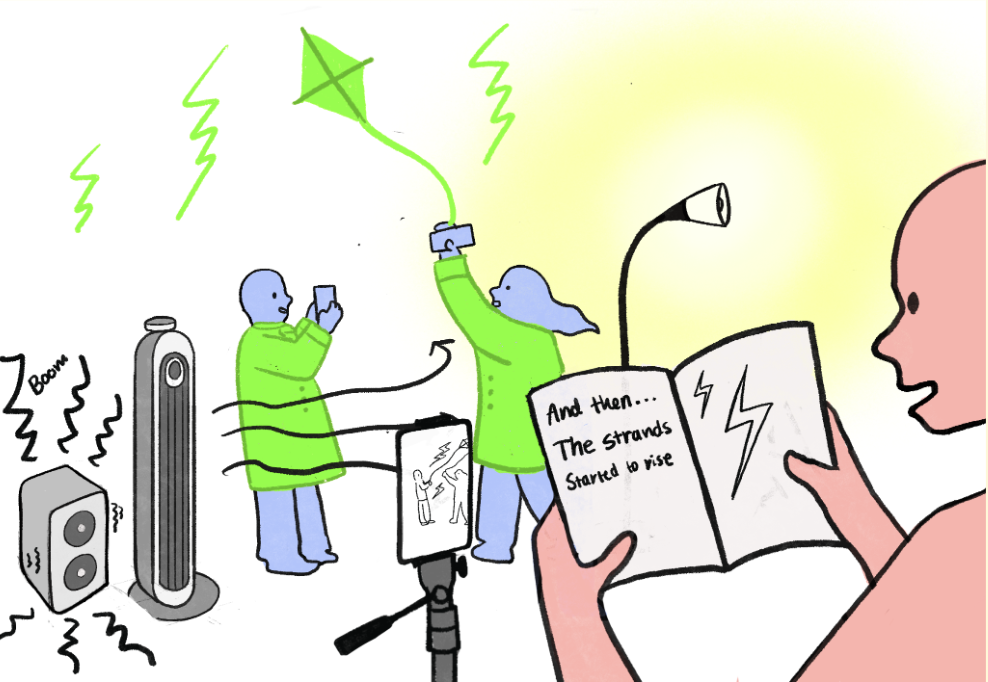}} 
  \hfill{}
  \subfloat[The Wind and the Sun.\label{fig:wind_and_the_sun}]{\includegraphics[clip,width=0.33\linewidth]{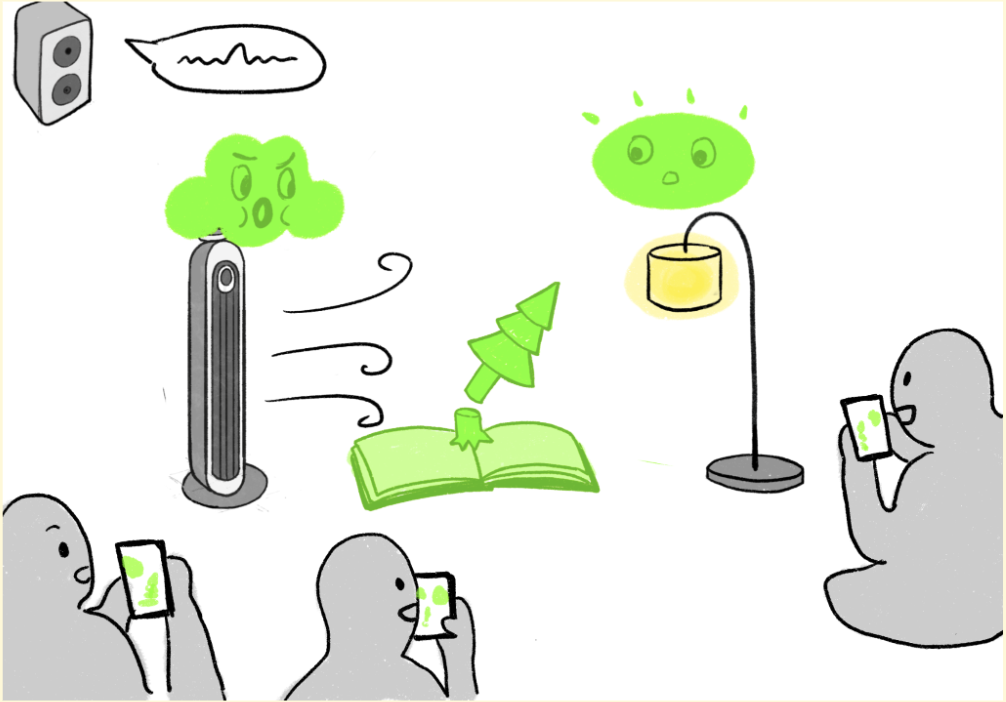}}
  \hfill{}
  \subfloat[Goodnight Moon.\label{fig:goodnight_moon}]{\includegraphics[clip,width=0.33\linewidth]{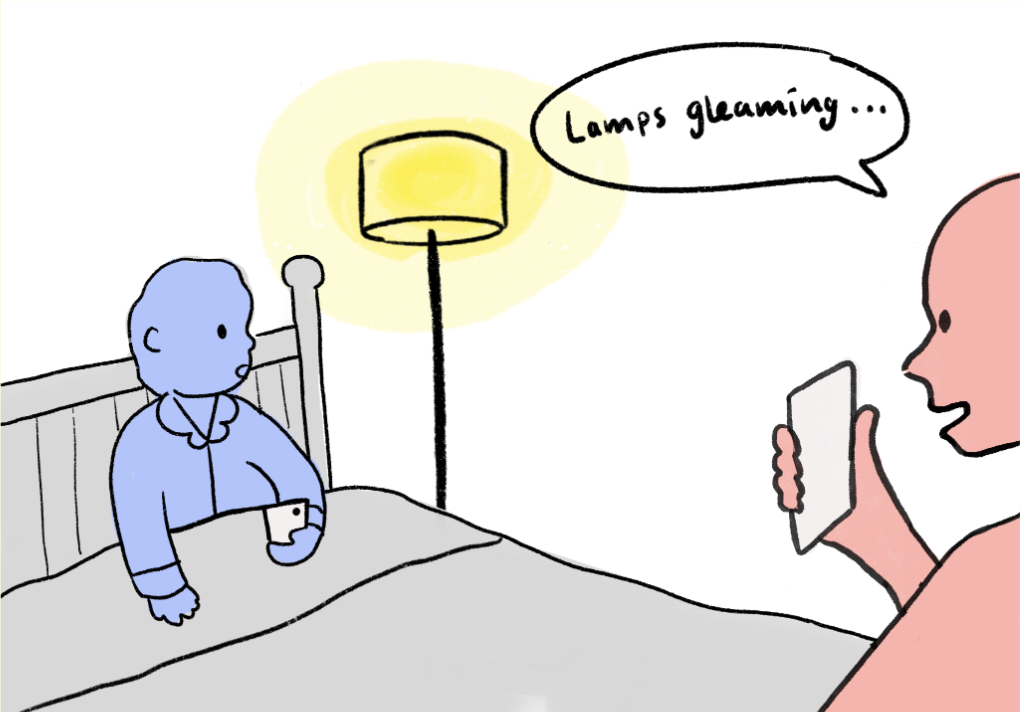}}
  \vspace{-.5pc}
  \caption{\lei{Three immersive stories designed via a mix of mobile AR and IoT devices. (a) The Benjamin Franklin story leverages IoT as background augmentation and enables users to actively participate as different roles in the story; (b) The Wind and the Sun story uses IoT as foreground augmentation and pictures users as the audience who watch the story proceeds; (c) The Goodnight Moon story uses IoT as foreground elements in the story, each of which responds to users' greeting.}}
  \Description{Three immersive stories designed via a mix of mobile AR and IoT devices. (a) The Benjamin Franklin story leverages IoT as background augmentation and enables users to actively participate as different roles in the story; (b) The Wind and the Sun story uses IoT as foreground augmentation and pictures users as the audience who watch the story proceeds; (c) The Goodnight Moon story uses IoT as foreground elements in the story, each of which responds to users’ greeting.}
  \label{fig:immersive_stories}
\end{figure*}

Inspired by the style of comic strips in storytelling and slide decks in presentation, we represent each immersive story using a sequential number of scenes.
The story proceeds as the system transitions from the previous to the next scene.
The transition of scenes can be triggered by either keyword detection (e.g., when the user says a keyword) or touch interaction (e.g., when the user's hand collides with an AR object in the physical world), similar to how the transition in Google slides is triggered by pressing a button on the mouse.
Each scene contains states of AR content (e.g., appearing/disappearing of a 3D model) and IoT devices (e.g., turning on/off a smart device).
We use three types of off-the-shelf IoT devices in these immersive stories: smart lights, smart fans, and smart speakers.
In the below sections, we present the design considerations of each immersive story.

\subsubsection{Benjamin Franklin Kite Experiment}
The Benjamin Franklin Kite Experiment is a story of Benjamin Franklin and his son taking a kite out during a storm to see if a key attached to the string would draw an electric charge (as seen in Fig. \ref{fig:ben_franklin}).
To play the Benjamin Franklin story, three users are required to be co-located with each user holding a phone.
At the beginning of the experience, users coordinate according to the app and assign each user a different role: a narrator, an actor of Benjamin Franklin, and an actor of Benjamin Franklin Jr.
After assigning the roles, users can see the two actors' bodies augmented with virtual costumes on them.
The app then asks the narrator to narrate the story by reading a physical book out loud.
As the narrator reads the story on the physical book, the app uses voice recognition to transcribe what the narrator says and automatically detect keywords in the transcription.
A list of keywords that are selected based on the physical book and their corresponding scenes are predefined in the app.
As one keyword is detected, the app will jump to its corresponding scene and wait for the next trigger.
In this story, we defined a total of six pairs of triggers and scenes, with the last trigger being a touch event between users' hands and a virtual key in AR.

In this experience, we focused on using IoT devices as background augmentation, having human actors interacting with the story, and having a narrator reading from a physical book.
The augmentation is considered as in the background since IoT devices add to the story's ambience, such as dimming the lights to simulate cloudy weather.
The visualization of costumes and interaction of touching the key are designed to facilitate the sense of participation.
Lastly, the design of reading a physical book is to increase the sense of physicality, where the physical interactions of flipping pages and reading printed text can be meaningful in the experience.


\subsubsection{The Wind and The Sun}
The Wind and the Sun story is from the famous Aesop's Fables, where the two characters quarrel about which of them is stronger (as seen in Fig. \ref{fig:wind_and_the_sun}).
Any number of users can play the story with each user holding a phone.
The system narrates the story through the smart speaker.
Users can tap on the screen to make the story proceed.
In this story, the IoT devices act as the driving actors of the story with visual augmentations from the phone's AR, where the system superimposes the Wind on the smart fan and the Sun on the smart light. 

Compared to the previous story, this experience is similar to a 4D movie with less focus on interactivity. The users take a passive role, and the primary user who would have been the narrator before now only has control of moving to the subsequent scene. 

\subsubsection{Goodnight Moon}

We recreated Goodnight Moon \cite{goodnightmoon}, an American children's story written by Margaret Wise Brown in 1947. The narrator greets objects around the room (e.g., ``Goodnight, red balloon. Goodnight lamp. Goodnight fan''). The greeting causes an appropriate change in the greeted AR or IoT device (e.g., the balloon goes down; the smart lamp turns off; and the smart fan stops blowing wind). The story progresses with keyword detection as the narrator speaks, similar to the Benjamin Franklin story. However in this experience, the story is shown on the screen rather than on a physical book (as seen in Fig. \ref{fig:goodnight_moon}).

Because the narrator explicitly greets the IoT devices, the devices take on more of a foreground role. While the audience still has a passive role, the narrator is now an active, interactive participant in driving the story. The one-to-one mapping of each narration line with an effect on some object (AR or IoT) creates a simple but magical experience.

\subsubsection{Implementation}
\label{system-implementation}
We implemented the three stories using Lens Studio\footnote{https://ar.snap.com/lens-studio}, which provides AR tracking, multi-user co-located experience, body pose detection and speech-to-text/text-to-speech capabilities. \dk{These features are necessary to enable the rich set of interaction techniques of the system, and motivated our choice for Lens Studio.} We connect and interact with the smart light using Kasa Smart's API \cite{kasasmart}. We use an IR-controlled fan, along with Bond Bridge \cite{Bond_2023} for networked control of IR devices. Lastly, we use Amazon's Echo smart speaker \cite{echoamazon} connected to the narrator's phone via Bluetooth.

\subsection{Jigsaw's Authoring Tool}
Following the \textit{trigger-and-scene} framework that was used to build different immersive stories, we introduce the authoring tool of Jigsaw which allows novices with little to no technical background to create these immersive stories with mobile AR and IoT devices.
We first present a system walkthrough of \lei{a step-by-step implementation of the Goodnight Moon story using Jigsaw's authoring tool}.
Then we introduce \lei{the underlying techniques including} our scene-based authoring approach and in-situ editing in AR.
Lastly, we demonstrate the generalizability of our authoring system by \lei{illustrating how to replicate the other stories, i.e., Bejamin Franklin and the Wind and the Sun.}

\begin{figure*}[t]
    \centering
    \includegraphics[width=\linewidth]{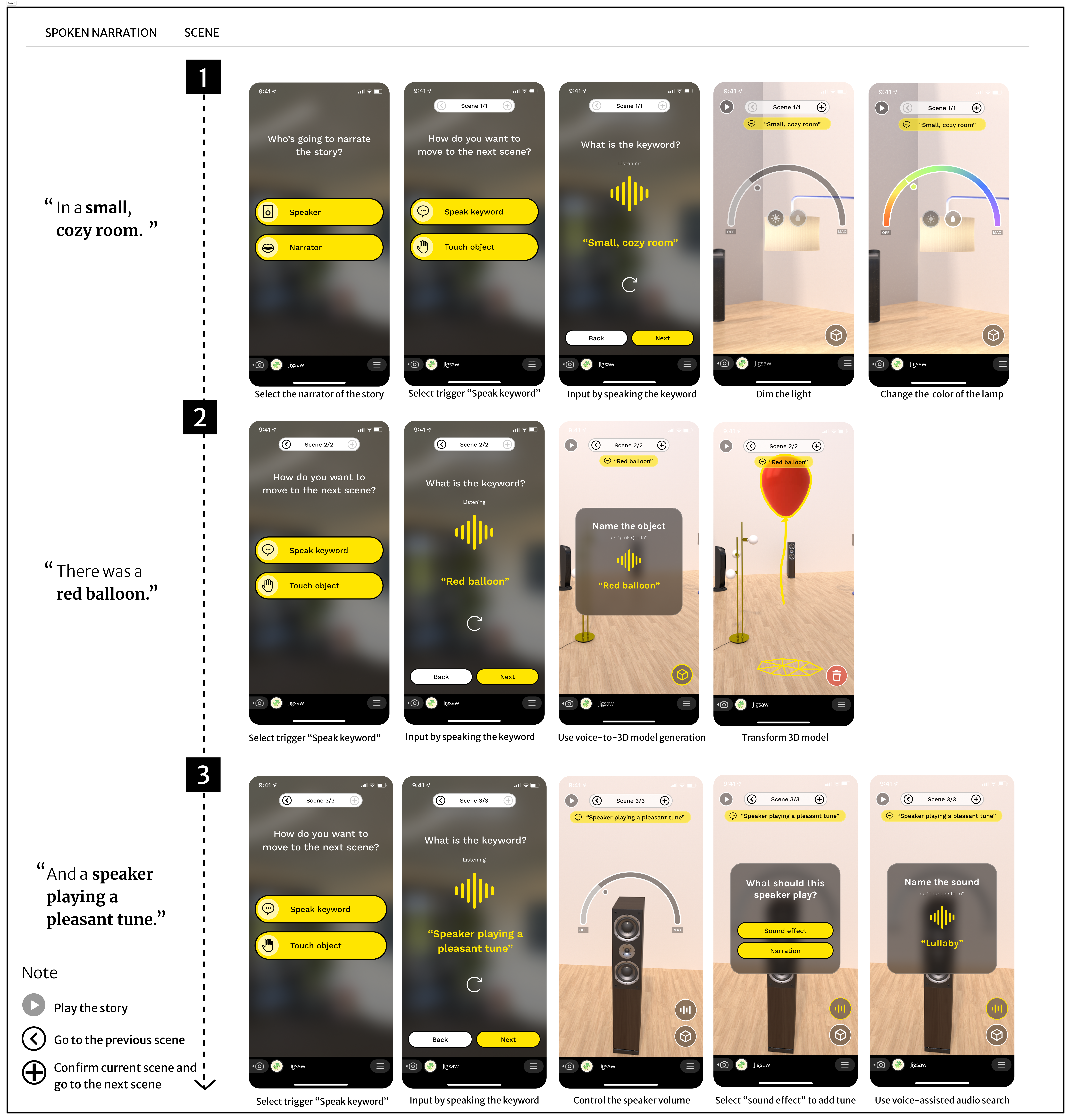}
    \caption{\lei{A step-by-step demonstration of creating the first three scenes of the Goodnight Moon story using Jigsaw's authoring tool.}}
    \Description{A step-by-step demonstration of creating the first three scenes of the Goodnight Moon story using Jigsaw’s authoring tool.}
    \label{fig:goodnight-moon-authoring}
\end{figure*}

\subsubsection{Scene-based Authoring and In-situ Editing}
Jigsaw enables novices to author immersive stories by creating a sequence of scenes in the editor, which draws inspiration from prior work on scene-based authoring (e.g., ~\cite{scenear2021}) and presentation tools such as Google Slides. 

\lei{In Jigsaw, users are able to define the following primitives of immersive stories via Jigsaw's authoring tool:
\begin{itemize}
    \item \textbf{Behaviors}: A behavior represents the state of either a virtual object in AR or an IoT device (e.g., appearing/disappearing of a 3D model, on/off of a smart light).
    \item \textbf{Scenes}: A scene represents the state of the immersive story and can consist of zero to multiple behaviors.
    \item \textbf{Triggers}: A trigger is used as the transition from the current scene to the next scene. Each scene progresses to the next through exactly one trigger that the user defines. 
\end{itemize}}

In Jigsaw, three trigger types are available: tapping on the screen, keyword recognition via speech, and touching a virtual object in AR.
Keyword triggers, as those in the Goodnight Moon story, allows users to define custom keywords for entering each scene.
This is incorporated since verbal narration has been a key component of storytelling experiences.
The touching trigger enables users to enter a new scene when their hands collide with an AR object.
This is one of the basic interaction in AR applications in order to add to the interactivity of the stories.
Lastly, tapping is the default trigger for stories \lei{that are narrated by the system} where the user enters a new scene by tapping on the screen.

In addition to defining \textit{triggers} to different scenes, Jigsaw allows novices to edit \textit{behaviors} of AR content and IoT devices within each scene directly in AR.
This is aligned with the spirit of What-You-See-Is-What-You-Get that numerous authoring tools employ to lower the barrier for end-users to create immersive content~\cite{flowmatic, ivy}.
\lei{An interface for importing 3D assets is shown on the screen when the user edits the scene.
By saying the name of the 3D asset that they want, users can add the 3D asset to the AR scene.
They can then adjust the placement of the asset by tapping and dragging the asset on the screen; they can also adjust the depth dimension by moving the phone in the physical environment while dragging the 3D asset.}
Jigsaw \lei{also enables} in-situ editing of IoT devices including smart lights, smart fans, and smart speakers.
A corresponding editing interface is popped up when the distance between the user's phone and the IoT device is below a pre-defined proximity threshold.
For the smart light, users can edit its color, brightness, and special effect.
The color is visualized in a semi-circular palette and the user can change its color by tilting the phone.
Users can change the brightness using the same control, where the value of 0 will turn off the light.
Lastly, users can add special effect of smart lights by saying the name (e.g. flickering).
We envision delegating the low-level implementation of special effects to advanced developers and only expose the effect name to novice creators.
Similarly, users can edit the on/off and intensity of the smart fan, and the on/off, volume, and special sound effects of the smart speaker.

\subsubsection{System Walkthrough: Building the Goodnight Moon Story}
We present a system walkthrough involving a general end-user who has little to no technical background and is motivated to use Jigsaw to create the Goodnight Moon story that end-users can play with their families and friends (as seen in Fig. \ref{fig:goodnight-moon-authoring}).

\lei{Opening Jigsaw's editor, the user first selects the narrator of the story to be users rather than the system.
After selecting the narrator, the user then enters the initial scene which is an empty scene.
The first scene that the user is going to build for the story is dimming and changing the color of the smart lights when the user says ``In a small, cozy room...''
From the initial scene, the user can create a new scene.
Before entering the new scene, the system prompts the user to define a keyword; in this case, the user speaks out the keyword ``small" and confirms.
After the trigger is defined, the user can edit the new scene by defining behaviors of the smart light. 
As the user walks closer to one of the lights, an editing interface of the smart lights is shown in AR based on the user's proximity to the lights and allows the user to edit the status of the lights.
The user can toggle between changing different attributes of the smart light including the brightness and the color.
They can then turn down the brightness and change the color of the smart light.
Jigsaw allows users to define multiple behaviors in a scene (e.g., they can walk up to other IoT devices to edit additional behaviors) though in this specific story the scene only contains the behavior of one device, i.e. the smart lamp.
In this way, the first pair of trigger (i.e., the keyword ``small, cozy room'') and scene (i.e., changing the brightness and color of smart lights) is recorded in the system.}

\lei{Upon finishing the first scene, the user can add a new scene, which saves the existing scene and jumps to the definition of a new trigger.
The second scene that the user is going to build for the story is making a red balloon appear when the user says ``There was a red balloon.''
The user then defines the corresponding keyword as the trigger and starts editing the new scene. 
In addition to changing the status (e.g., on/off and intensity) of IoT devices, Jigsaw also allows users to create, manipulate, and delete 3D models or animation.
In this case, to add a balloon model, the user can tap on the corresponding button and say the model name (e.g., ``red balloon") directly to the system.
The system then dictates the model name spoken by the user and shows the corresponding model in AR.
The user then places the content in the physical space by simple drag-and-drop interaction; alternatively, they can delete it by dropping it to the trash icon on the screen.
The system can create a fade-in, fade-out, and moving animation by default.
For example, when scene 1 does not contain a balloon and scene 2 has one, the system can create a fade-in effect for the balloon in scene 2 during playback.
Similarly, the system automatically applies a translation animation to a 3D object by moving from the position of the previous scene, to the position of the next scene during playback.
In this way, the user can author scenes such as a moon model appears and disappears.
After creating the two scenes, the user can create the following scenes in a similar manner (note that the Goodnight Moon story contains 11 pairs of triggers and scenes in total) as the story proceeds.
Jigsaw allows users to navigate and edit previous scenes.
For example, the user can navigate to the first scene and change the color of the lamp.
Finally, using Jigsaw, the user can play the storytelling experience immediately in the authoring environment.
}

\begin{figure*}[t!]
  \centering
  \subfloat[Adding narration.\label{fig:add_narration}]{\includegraphics[clip,width=0.24\linewidth]{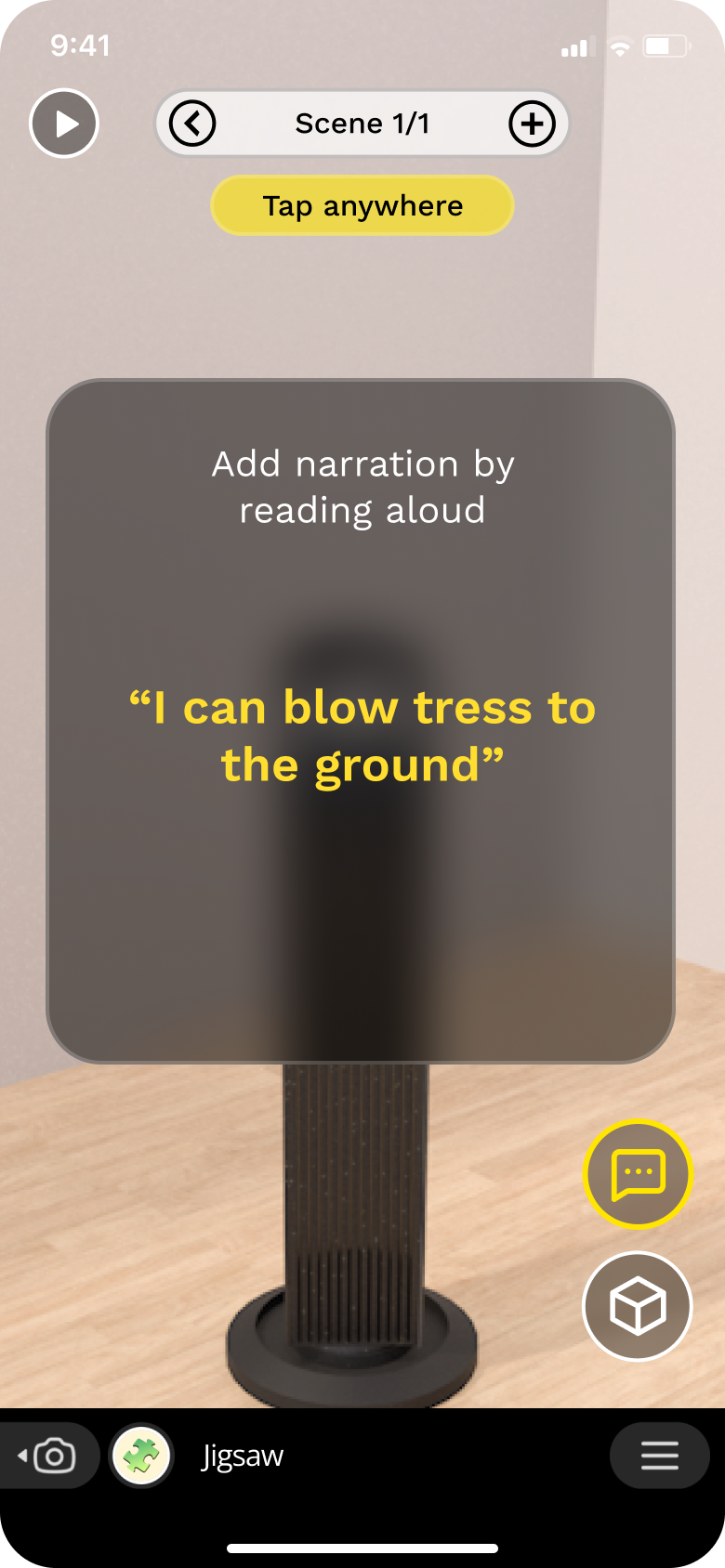}} 
  \hfill{}
  \subfloat[Editing the intensity of the smart fan.\label{fig:adjust_fan}]{\includegraphics[clip,width=0.24\linewidth]{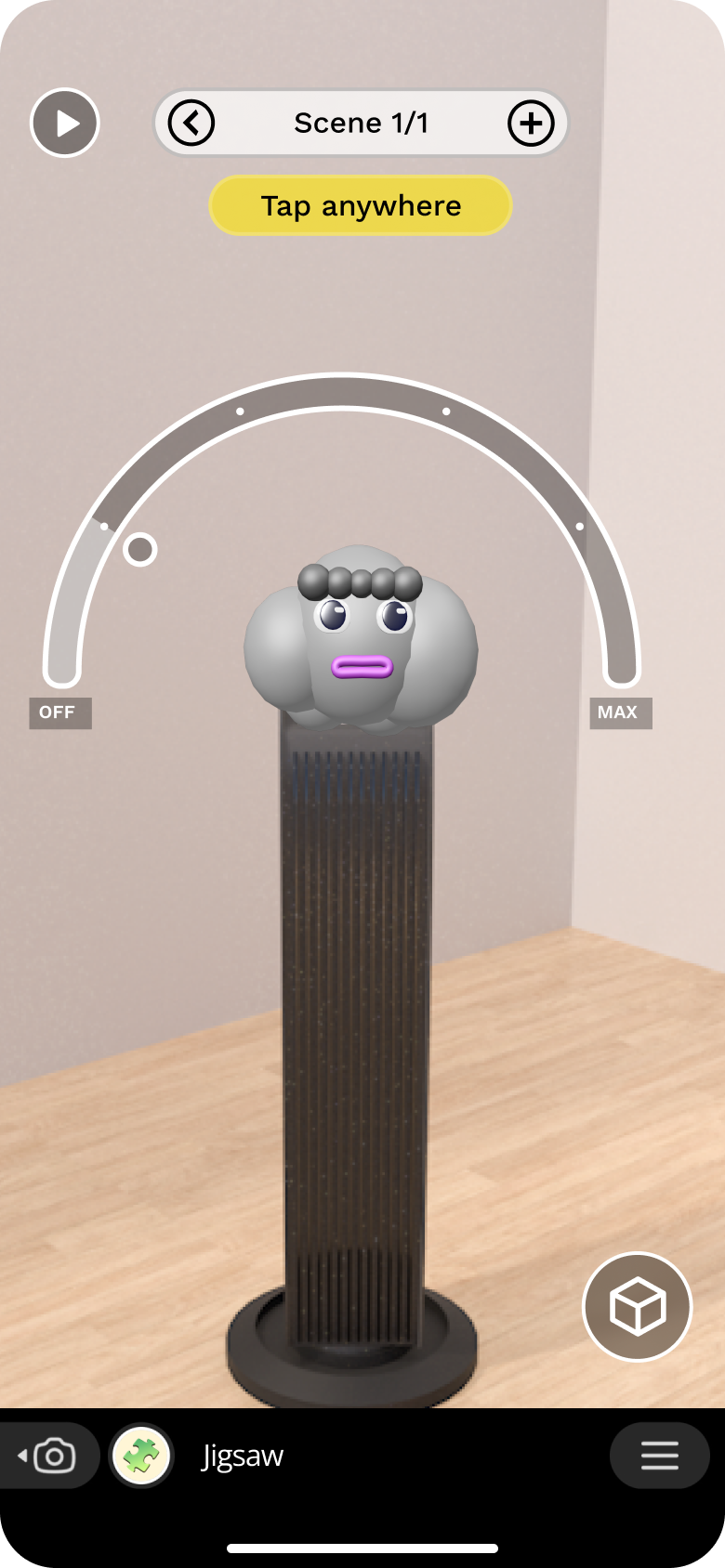}}
  \hfill{}
  \subfloat[Defining the object for touch interactions.\label{fig:touch_interaction}]{\includegraphics[clip,width=0.24\linewidth]{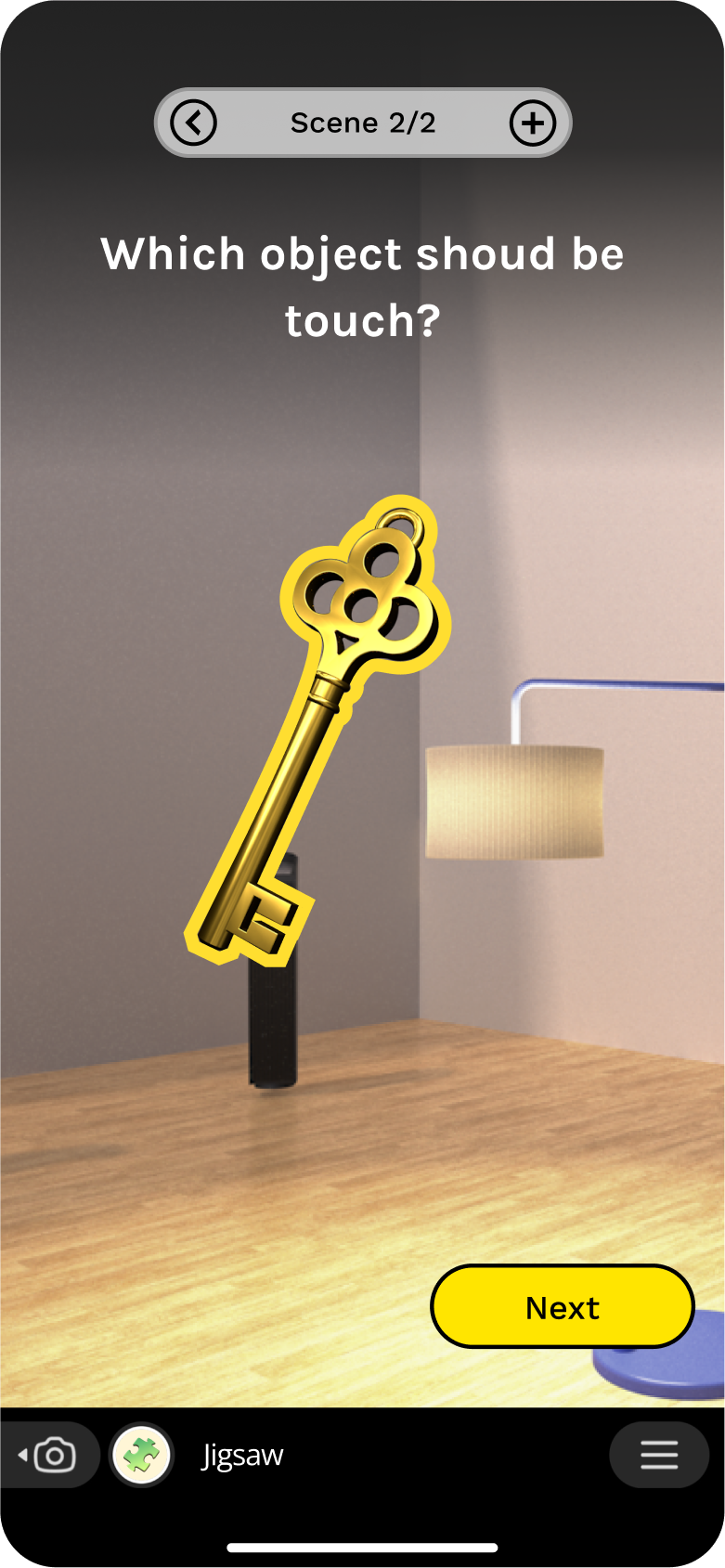}}
  \hfill{}
  \subfloat[Selecting pre-defined effects of models.\label{fig:key_effect}]{\includegraphics[clip,width=0.24\linewidth]{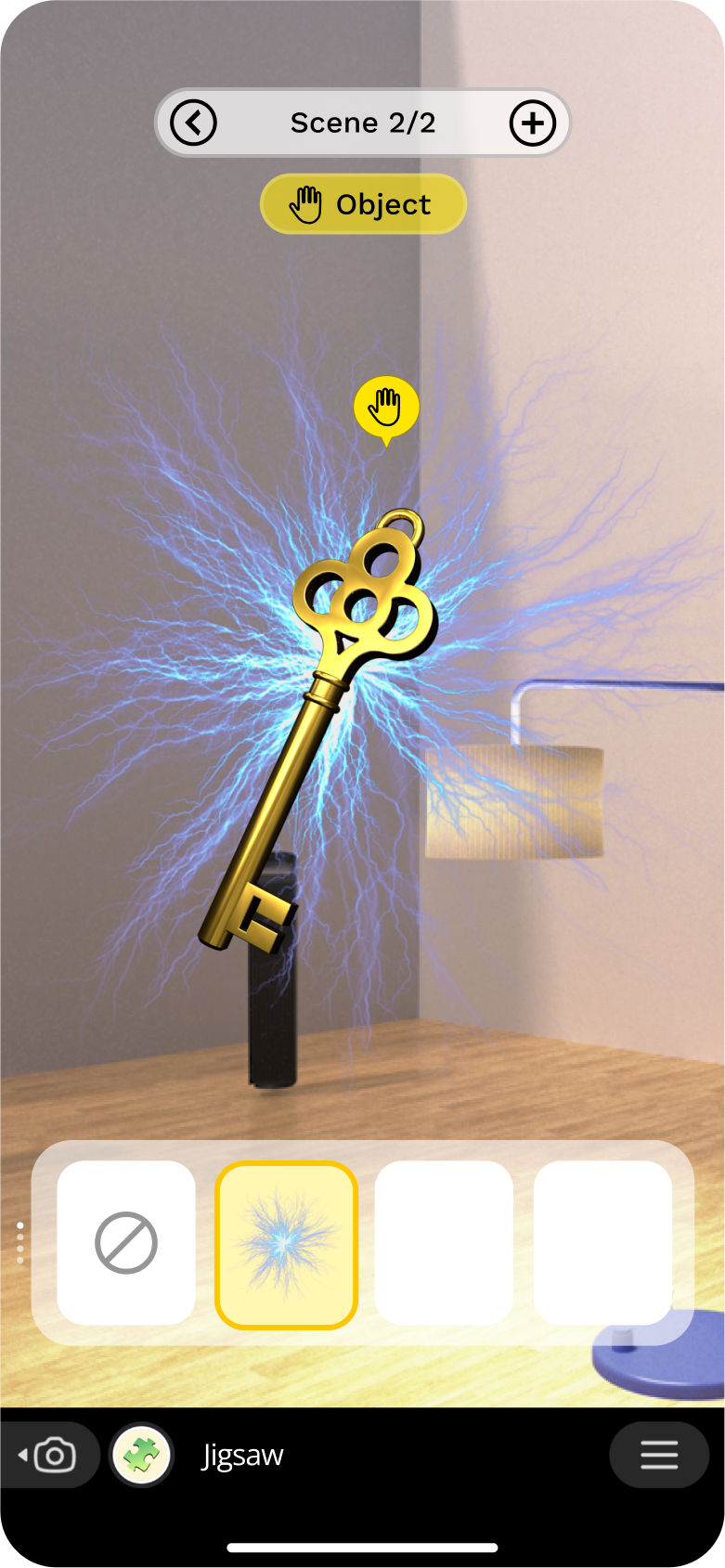}}
  \caption{\lei{Authoring interfaces for adding narration, editing smart fans, and defining touch interactions. (a) For self-narrated stories like the Wind and the Sun, the default trigger for each scene is tapping on the screen and users can add narration to each scene. (b) Users can edit the intensity of the smart fan with the Wind character anchored to it in The Wind and the Sun story. (c) When defining the touch interaction in the Benjamin Franklin story, users can select which object to touch by tapping on the object. (d) Users can also choose built-in effects of 3D models when the touch interaction is triggered.}}
  \Description{Authoring interfaces for adding narration, editing smart fans, and defining touch interactions. (a) For self-narrated stories like the Wind and the Sun, the default trigger for each scene is tapping on the screen and users can add narration to each scene. (b) Users can edit the intensity of the smart fan. (c) When defining the touch interaction in the Benjamin Franklin story, users can select which object to touch by tapping on the object. (d) Users can also choose built-in effects of 3D models when the touch interaction is triggered.}
  \label{fig:authoring}
  \vspace{-.5pc}
\end{figure*}


\subsubsection{Generalizability of the Authoring System}
A key factor that makes our authoring system generalizable is that the the triggers and scenes that users create are open-ended and not coupled with any specific story.
For example, a user can define any keyword or touching with any AR object as the trigger and use the brightness of smart lights for any kinds of story.
To demonstrate the generalizability of the authoring system, we use replicated examples~\cite{ledo2018evaluation} by replicating the other two immersive stories using the same system.

To replicate the Benjamin Franklin story, users can define pairs of triggers and scenes like they do in creating the Goodnight Moon story.
\lei{The unique challenges of creating this particular story include allowing multiple behaviors in one scene and defining touch triggers.
Different from the Goodnight Moon story, the Benjamin Franklin contains multiple behaviors triggered by one keyword in a scene.
For example, when the user says ``when the clouds first passed over'', the smart fan will turn on and the AR cloud model will appear.
To achieve that, users can walk up to the smart fan and adjust its state, and then import the 3D model of a cloud while staying in the same scene.
After editing all the behaviors in a scene, the user can proceed to add a new scene, which indicates finishing the current scene. 
}
The Benjamin Franklin story also incorporates touch triggers, where the user touches the virtual key model and triggers the flickering effect of smart lights, as seen in Fig. \ref{fig:touch_interaction} \& \ref{fig:key_effect}.
Jigsaw accomplishes this by allowing users to select trigger types \lei{besides keyword recognition} when adding a new scene.
After selecting touching as the trigger type, the user can then select the object for collision detection by touching it directly in AR.

The Wind and the Sun story is different from the Goodnight Moon example since it requires the system instead of the user to narrate the story.
Jigsaw enables this by allowing users to select narrators at the very beginning of the authoring experience, as seen at the beginning of Fig. \ref{fig:goodnight-moon-authoring}.
After selecting the system as the narrator, the user can add a narration to each scene by saying it directly to the system.
The system can then transcribe the narration and play it through the speaker during the playback, \lei{as seen in Fig. \ref{fig:add_narration}}.
\lei{When the system is assigned as the narrator, the system will omit trigger selection when adding a new scene and the trigger type goes to default, i.e. tapping on the screen.
Users can import 3D assets such as the Wind character and place it in the physical environment such as anchored to the smart fan.
They can also edit the intensity of the smart fan as seen in Fig. \ref{fig:adjust_fan}.}

\subsubsection{Implementation}
The authoring system is implemented using Lens Studio for displaying and interacting with AR content.
The connection with IoT devices is built using the same technique as the implementation of the immersive stories mentioned above.
We used the built-in voice recognition and hand tracking component in Lens Studio to enable users to define keyword and touch triggers.
The playback of the final experience among multiple users is implemented using the built-in connected lens component.
\section{Evaluation}
We conducted an in-lab user evaluation with 20 participants (10 groups) with two goals: \textit{(i) to investigate the benefits and challenges of immersive storytelling experiences with both virtual and physical augmentation, and \textit{(ii)} to assess the usability and utility of the authoring system}.

\subsection{\lei{Participants}}
We recruited participants from a technology company in the United States through Slack posts and e-mails.
Participants \dk{(ages 21 to 36)} were randomly paired into 10 groups and \lei{volunteered for the one-hour study}.
Our study was approved by the privacy \& legal team in our institution.
\lei{Participants had either strong ties (e.g. friends) or weak ties (e.g. colleagues) within each group. 
All but one of the participants had prior experience of using IoT devices such as Amazon Alexa.
All but two of the participants had used AR products such as mobile AR or AR glasses before the study.}

\subsection{\dk{Apparatus}}
The immersive experiences used three iPhone 13 Pro Max devices for AR visualizations and interactions. Same devices described in section \ref{system-implementation} are used to enable physical augmentations, such as lightning and storm effects.

\subsection{Study Procedure}

\begin{table*}[t!]
  \caption{The first three scenes of the Goodnight Moon story for the authoring task..}
  \label{tab:authoring}
  \begin{tabular}{lll}
    \toprule
    \textbf{Scenes} & \textbf{Keywords} & \textbf{Behaviors} \\
    \midrule
    Scene 1 & ``small, cozy room" & The brightness of the smart light becomes 20\%. \\
    Scene 2 & ``red balloon" & A red balloon model appears in the room. \\
    Scene 3 & ``speaker playing a pleasant tune" & The smart speaker plays the sound ``lullaby". \\
  \bottomrule
  \end{tabular}
  \vspace{-.5pc}
\end{table*}

Our study was divided into two sessions, each lasting 30 minutes.
Prior to the study, participants first signed an agreement form that confirms their consent for our data collection and usage.
They also completed a short entry survey about their demographic information and their prior experience of \lei{using creation tools for content such as image/video (e.g., Adobe Photoshop) and presentations (e.g., Google Slides)}.

\subsubsection{Experience session}
In the first session, we asked the participants to play the three immersive stories that we created in the order of Goodnight Moon, the Wind and the Sun, and Benjamin Franklin.
\lei{We started with Goodnight Moon for its simplicity, where each trigger links to one behavior, and concluded with the more complex Benjamin Franklin story involving multiple behaviors and interactions.
We chose this story order to progressively familiarize users with the immersive storytelling landscape, aligning with our goal of exploring this emerging paradigm of immersive stories.}
Before playing each story, participants were debriefed about the story background and their roles in the story.
Since the Benjamin Franklin story requires three people to experience, one member from our study team acted as the Benjamin Franklin Jr.
\lei{In this story, one participant acted as the narrator holding the physical book and the other participant acted as the character listening to the story and touching the key. 
Our team member filled this audience role their role was only listening to the story, which overlapped with existing participant roles, ensuring all participation aspects were experienced.}
After experiencing the three stories, we conducted a semi-structure interview with both participants to ask about their experiences and the benefits and challenges of the immersive stories.

\subsubsection{Authoring session}
In the second session, we spent the other half an hour evaluating the authoring system with one randomly picked participant from the group.
\lei{We decided to ask participants to first experience and then recreate as we believed it is essential for them to understand this new paradigm of immersive stories (i.e. ``what" immersive stories are) before coming up ways to craft them (i.e., ``how" to create immersive stories).
The participant was first given a 10-minute tutorial introducing the functionalities of Jigsaw's authoring tool, including defining triggers, creating scenes, and editing behaviors of 3D models and IoT devices.}
After ensuring the participant understood how the authoring system works, we then asked the participant to complete the task of recreating the Goodnight Moon story that they experienced in the first session.
\lei{We asked participants to replicate existing stories rather than creating their own since having one consistent story across all participants can help us determine whether participants were able to create immersive stories using Jigsaw’s authoring tool. 
In the evaluation of Jigsaw's authoring tool, we also emphasized the creation process rather than the story content.}
During the task, the participant was provided with the complete script on a piece of printed paper including keywords and their corresponding behaviors.
\lei{An example of the first three scenes in the script can be seen in Table \ref{tab:authoring}.}
We provided this so that the participant could focus on \textit{how} to create the story using the system instead of \textit{what} to create.
During the task, the study team was not allowed to provide any help unless the participant explicitly asked for help.
Any question that the participant asked during the task was noted down.
The task included setting up 11 pairs of keywords and behaviors and took 10 minutes in total.
Finally, we conducted a 10-minute semi-structure interview with the participant to ask about the usability and utility of the authoring system.

\subsection{\lei{Data Analysis}}
\lei{We aggregated the entry survey data and conducted a thematic analysis~\cite{braun2006using} of the transcriptions of the retrospective interviews for both the consumption and the creation of immersive stories via Jigsaw.
Two researchers first developed a codebook based on four randomly selected participants’ transcripts, including two from the consumption interview and two from the creation interview, identifying and labeling similarities across participants’ experiences with Jigsaw. 
The researchers discussed and refined the codebook on another subset of participant transcripts. 
We then validated the codebook on two additional participants’ transcripts, individually coding the same transcripts. 
After achieving high inter-rater reliability, with a Krippendorff’s $\alpha$ above 0.8, they divided and coded the rest of the transcripts in parallel.
Finally, we discussed and grouped related codes into themes according to our research questions.}
\section{Results}
Participants' experiences playing immersive stories during the study suggested that Jigsaw's storytelling experiences with both virtual and physical augmentation were immersive, engaging, and memorable.
\lei{In the evaluation of the authoring system, all participants were able to complete the creation task and applauded the system for being easy to use and generalizable for creating a wide variety of stories.}
We center our findings around the two evaluation goals, \textit{(i) benefits and challenges of immersive stories}, and \textit{(ii) the usability and utility of the authoring system}

\subsection{Benefits and Challenges of Immersive Stories}
\lei{Overall, our results suggest that the design of Jigsaw made the storytelling experience immersive, engaging, and memorable by adding augmentation of IoT devices.
Participants felt they had control over the storytelling experiences via various forms of participation and interactions with elements of the story.
Participants also highlighted the versatility of these immersive stories for various population and scenarios.
}

\subsubsection{IoT devices made the experience immersive, engaging, and memorable.}
\lei{Immersion is often associated \textit{extensiveness} (i.e., the range of sensory modalities accommodated) and \textit{surroundings} (i.e., the extent to which the experience is panoramic rather than limited to a narrow field) of the experience~\cite{slater1997framework}.
Our results suggested that, in Jigsaw, the augmentation via IoT devices increases the \textit{extensiveness} to make the storytelling experience more immersive by invoking more senses such as haptics from the smart fan.}

\begin{quote}
    ``if you just had AR simulated wind, there's something that's not as exciting about the fact of having real wind and having that effect come together...the wind being real because you can feel it."-P9B
\end{quote}

\lei{In line with our motivation to create AR experiences beyond just ``pixels" on the screen, participants also reported the storytelling experience being immersive as it extends from the screen to the IoT devices in the \textit{surroundings}.}
\begin{quote}
    ``it's not only are you seeing it on the screen, it had panned out a little bit further and not only do you see it on the screen, but you're feeling it too, so it's 4D as opposed to just the AR experience of seeing it.'' -P11B
\end{quote}



\lei{With the enhanced immersion, the storytelling experiences became memorable and engaging to participants due to the various neural stimuli involved in the experiences.}
\begin{quote}
    ``I think it (without IoT) would've been less of an immersive experience and also probably little less memorable. So you remember, "Okay, at some point the moon went to sleep, the lights went out," and it's a bit easier to remember visual cues than it is just reading it off. And then also staying with the story was a lot easier because things are changing around and keeping me engaged." -P2A
\end{quote}

\subsubsection{Users' participation and interaction improves the sense of immersion and agency.}
We also found that allowing users to act as different roles (e.g., narrators or characters in the stories) in the stories enhanced the sense of immersion.
This is because when users act as characters in the stories and their interaction with the story elements triggers certain effects, it \lei{reinforces} the impression that they are embodied as the character; when users act as narrators, they tend to focus more on the storyline and feel immersed.
\begin{quote}
    ``It makes you feel like you are in a theatre with the whole background noise, the light changes, and the fact that you can touch things virtually... I was very involved in everything going on, so I was definitely very immersed." -P5A
\end{quote}

\begin{quote}
    ``I think being the narrator helped me be immersed in the story more. I think maybe because I had to focus on the story more... Since I'm narrating, I have to think of what's going to happen next or think what I have to say next. Whereas when I'm an audience member, I think it's a little easier to get distracted by what's going on and then lose what the computer's saying." -P7A
\end{quote}

\lei{Agency refers to ``the satisfying power to take meaningful action and see the results of
our decisions and choices"~\cite{murray2017hamlet}. 
In addition to prior digital storytelling experiences that explore the sense of agency via questions and answers~\cite{storybuddy2022}, we found that being the narrator in the immersive stories can give participants the sense of agency since they feel that they have control over how the story goes, by closely examining the mapping between triggers and scenes based on their own paces.}
\begin{quote}
    ``I definitely enjoyed being a narrator more because I felt more in control of the actions that I was committing throughout the story. So let's say I was saying the story and then something came on the AR screen, I was able to look around and say, okay, I'm done looking at that. I'm ready for the next thing." -P7A
\end{quote}

\subsubsection{Sensory overload is the primary challenge in immersive stories.}
While Jigsaw's immersive stories are immersive, engaging, and memorable, we found that our design of immersive stories also comes with the challenge of sensory overload.
\lei{We found that participants' attention was occupied by participation as different roles, AR effects on the phone, and IoT effects in the surroundings, making it difficult for them to catch up with all the changes.
For example, participants reported missing the IoT behaviors easily when they were focusing on the AR effects on the phone.}

\begin{quote}
    ``I think if the experience doesn't happen at the same time... that would make sense, because I feel like if you're already dealing with something flashing in your phone and you're going to blink and miss what's in your world, because you're really focusing on what's on your phone there." -P11A
\end{quote}

\lei{In addition to the tension between AR content on the screen and IoT behaviors in the surroundings, participants also expressed sensory overload of both reading the physical book and paying attention to the changes on the phone.} 
\begin{quote}
    ``I love reading books. I love reading to children. I didn't need to have a phone. The reader in that case doesn't need to have any devices. I don't need to be a part of the immersive experience. I can just be the reader reading to the people in the room. So I didn't really like having the phone sitting beside me because I'm just like, now I'm tempted to pick up the phone, but I'm trying to read this story in a really nice way. So it just felt like there was this weird divide and I felt like I was being pulled in two different directions and not really being able to fully be in the moment." -P4B
\end{quote}

\subsubsection{Immersive stories are versatile for various population and scenarios.}
Lastly, we found that Jigsaw's immersive stories are versatile for both ambient and interactive scenarios, and for both kids and adults.
\lei{This corresponds to our motivated design explorations of using IoT devices for both foreground and background augmentation (Table \ref{tab:design}).
For example, participants commented that the foreground augmentation of IoT devices could be used for stimulating children to keep them engaged and for offering interactive gaming experiences such as escape rooms for adults.}

\begin{quote}
``because children have low attention spans, they're all looking at screens all day long, watching those 15-second unboxing YouTube videos... they need to be stimulated to be engaged, so that's why I see this being working wonders with children" -P4A
\end{quote}

\begin{quote}
    ``I can't tell you of how many escape rooms I know where you trigger one thing happen in the room, like flickering lights. That would be perfect for it." -P11B
\end{quote}


\lei{In addition, we found that when Jigsaw's immersive stories can also be used for ambient scenarios, when IoT devices were used as background augmentation to set the mood of the experience.}
\begin{quote}
    ``I think one thing that I would love to see, so maybe for the adult story part, would be more horror type of stories. I could see a potential there because you have lights flickering, you have sort of sounds that come out of nowhere or wind that starts blowing." -P8A
\end{quote}

\begin{quote}
    ``I have a set bedtime and 30 minutes before that I'm supposed to do this end of the day meditation, where I watch a sunset. And I just thought it reminded me of that, because it would be really cool to see that in AR and then maybe you power down your devices, because I'm supposed to be off my phone until bedtime. You power down your TV and electronics and screens and stuff, lower the lights... that type of transition." -P10A
\end{quote}

\subsection{Usability and Utility of the Authoring System}
\lei{We also sought to understand the usability and utility of Jigsaw's authoring tool, as we are curious whether users are able to create these new immersive stories. 
We found that the design of Jigsaw's authoring tool has a low entry of barrier and is generalizable due to the open-endedness of the triggers.
However, the Jigsaw's authoring tools face challenges of creating more complex stories such as creating custom AR animation, coming up with triggers and behaviors, and authoring longer stories.
We describe these in detail below.}

\subsubsection{The authoring system has a low barrier of entry.}
Participants commented that the authoring system is easy to learn, partly because the in-situ interface that Jigsaw's authoring tool offers allows users to directly edit behaviors of AR and IoT devices in AR with little technical literacy such as imperative programming.
\begin{quote}
    ``I think if I was a non-technical person, it was pretty easy to use because you just point at an object and then you can interact with it. So that mechanic, rather than me specifying things through a complicated user interface or a bunch of text, makes it a lot easier to just as a lay person to be, "Oh, I want the lights to turn on or to be red," and then you can just do it...I've actually made a lens that controls Internet of Things before, but I had to do all of this through script and code, and it's not simple to use in that sense." -P8A
\end{quote}

\lei{This is in line with the benefits of prior authoring tools for immersive experiences that take advantage of directness (e.g., ~\cite{lee2004immersive, flowmatic}). 
In addition, participants reported that the authoring system makes it available to settings other than theaters due to less production cost to create immersive storytelling experiences.}
\begin{quote}
    ``For me, I almost immediately thought of preschools and kindergartens... especially at younger ages, you don't have that much production... so, just having, I would say, devices take cares of a lot of that."-P2A
\end{quote}

\subsubsection{Open-ended triggers make the system generalizable and fun to use.}
\lei{Participants commented that the keyword triggers were open-ended, which made the authoring tool generalizable and fun to use.
The ability to define custom keyword triggers, powered by Jigsaw's speech recognition capabilities, made users feel like they could ``do anything with it" (P10B).}
\begin{quote}
    ``I guess the benefits are with the keyword part that's super open-ended versus just on a Google Slides, you know, have only preset options, and so you're somewhat limited to the options that they provide you with. But the keyword element being something where you could just really make it any word that you want to say, it makes that part fun for the interactions themselves"-P9A
\end{quote}

The flexibility to specify any number and any types of behaviors triggered also enhanced the open-ended nature of the triggers.
\begin{quote}
    ``It's actually a story so that I can see this stuff can put me into a situation where I can probably trigger multiple things following the story as long as the story is intuitive and easy to follow. So that's definitely a lot add on there." -P5A
\end{quote}

\begin{quote}
    ``It would really be able to do a whole lot for each keyword or throughout the entire story that it's really going through a full experience. I like that it's not just one thing, so not just one keyword equals one action. One keyword can equal ten actions if you have it set up correctly." -P3A
\end{quote}

\subsubsection{Creating complex stories is challenging using Jigsaw.}
We also found that the expressiveness of our authoring system is limited when participants wanted to author special animation and more complex behaviors of objects such as making a shaking animation or setting the duration for the status of objects' behaviors.
Authoring these behaviors typically requires being able to edit lower-level attributes of the experience such as time, mesh, and position.
\begin{quote}
    ``Maybe if you wanted to make it feel like the ground was shaking or something like that, that would be hard."-P9A
\end{quote}

\begin{quote}
``If you wanted to have a flicker or setting a duration of the fan so it's like, oh, it's on for 10 [seconds] off for 10 [seconds], back on for another 30 [seconds]. Something like that, having more complex actions for each object would be really interesting." -P3A
\end{quote}

\subsubsection{\lei{Authoring long stories is challenging due to the limit of creation support.}}
Finally, we found that Jigsaw's authoring tool is suitable for authoring short stories rather than long stories due to the lack of support for duplicating triggers, scenes, or behaviors, and for navigating longer sequence of scenes.

\begin{quote}
    ``it would be too much work to copy from one scene to the next and let's say you want to have duplicated scenes or something like that. But yeah, I feel like children stories because they're kind of simple to create." -P7A
\end{quote}

\begin{quote}
    ``I think this one, like I said, was a bit more tedious because editing on for example, Adobe Premier Pro, I was able to just pick exactly what parts I want and put them next to each other whereas with this one, it's based on the keyword's is the next line. The more I have, the more work it's going to be, whereas the more clips I have might not necessarily be a linear growth at the time." -P2A
\end{quote}


Participants also commented that they might encounter challenges of coming up with pairs of triggers and scenes if they were going to create their own stories and would like more creation support during this process.

\begin{quote}
    ``I feel like it should have said something like, "Okay, now do you want to trigger the lights, the audio, or the fan?" and then you could decide. You already gave me the script and the script told me exactly what to do, but if I was trying to just create my own story from scratch for the very first time, at least once when you first make your first story there should be more prompts." -P4A
\end{quote}

\section{Discussion}

\lei{Overall, we found that Jigsaw enables immersive, engaging, and memorable storytelling experience through the combination of mobile AR and IoT devices.
We found the augmentation via IoT devices made the stories more immersive by adding more senses to the stories and expanding beyond ``pixels" on the screen.
Users' participation that Jigsaw enables in the immersive stories also enhances users the sense of immersion and agency.
In addition, our findings suggest that immersive stories should be carefully designed as mobile AR, IoT effects, and participation can cause sensory overload.
We found that the immersive stories are versatile for various population and scenarios.
Our results also revealed that participants were able to use Jigsaw's authoring tool to create immersive stories.
Our authoring tool can also enable users to create diverse stories due to the open-endedness of the keyword triggers.
However, our authoring tool has challenges in expressiveness for creating complex behaviors of objects or effects.
We highlight the opportunities and challenges of immersive storytelling experiences using mobile AR and IoT devices, and discuss design implications and limitations based on our results.}

\subsection{Design Implications and Future Directions}
\lei{Based on the results of evaluating Jigsaw's immersive stories and authoring tool, we highlight the opportunities and challenges of both consuming and creating immersive storytelling experiences via AR and IoT. 
We focus our discussions on how to make digital storytelling experiences more immersive via physical augmentation and user participation, how to create immersive storytelling experiences by enabling open-ended triggers and scenes, and how to support authoring more complex behaviors of AR and IoT.}

\subsubsection{Adding immersion via physical augmentation and user participation}
We found that the augmentation of IoT devices and user participation enhanced the sense of immersion during storytelling experiences.
This \lei{extends} prior effort of bringing in physical augmentation and participation for immersive experiences such as 4-D theatres and participatory plays~\cite{spacemountain}.
Our study indicates that enhancing readily available IoT devices, instead of using specialized, costly equipment, can effectively promote \lei{immersive storytelling experiences}.
\lei{Our results also align with prior exploration of dimensions of immersion including \textit{extensiveness} and \textit{surroundings}~\cite{slater1997framework}.
We found that Jigsaw's immersive stories can effectively increase the \textit{extensivenss} of immersion by adding in sensory augmentations and the \textit{surroungdings} by extending the storytelling experience beyond just the content on the screen.}
We encourage future researchers and practitioners exploring immersive stories integrate elements of physical augmentation and user participation to add to the immersion of stories.
\lei{For example, future immersive stories might explore how to extend the experience beyond visual displays on the scree via more IoT devices such as rombots and projectors which can be used for diverse tasks and content.
They may also explore immersive stories that involves more human senses such as haptics~\cite{abtahi2018visuo} and tastes~\cite{brooks2023taste}.}
They can also explore more means of participation beside being narrators, actors, and audience and investigate how they can affect users' sense of immersion during storytelling experiences.
\lei{For example, prior work has explored the involvement of generative AI for storytelling experiences~\cite{chung2022talebrush, storybuddy2022} and future research can explore how these new roles of user collaboration with AI could affect their sense of immersion in storytelling experiences.
In addition, another key insight that we gained from the results is that participants faced challenges of catching up with everything in the immersive stories due to sensory overload.
Therefore, future research should explore how to design immersive storytelling experiences that could balance various sensory augmentations and modalities of participation.}

\subsubsection{Authoring immersive stories via open-ended triggers and behaviors}
Our findings show that Jigsaw's open-ended approach to trigger and behavior definition enables it to be widely adaptable for authoring diverse stories. Users can assign a single keyword (e.g. "balloon") or behavior (e.g. smart light brightness change) to serve different narrative functions across multiple stories. 
We found that this flexibility can empower participants to get creative and develop novel ways of linking triggers with behaviors for diverse immersive experiences.
\lei{This aligns with prior work in creativity support tools that aims to enable people to concentrate on creative designs by removing low-level implementation and supplying appropriate building blocks~\cite{greenberg2007toolkits}.
We suggest future research and design in authoring immersive storytelling to consider a range of triggers and behaviors that are essential and open-ended for building immersive stories.
For example, in the context of authoring context-aware IoT applications, prior research has explored triggers such as the user's location and the state of IoT devices to activate events of IoT devices~\cite{wang2020capturar}.}
While being intuitive and versatile for authoring immersive stories, Jigsaw's authoring tool also has challenges of authoring long stories.
Participants also reported challenges coming up with the appropriate triggers and their corresponding behaviors.
Future research could thus explore ways to automate the process of defining triggers, scenes, and behaviors.
For example, one can imagine narrating the story verbally for the first time and the system could intelligently detect and record triggers and behaviors on the fly.

\subsubsection{Supporting end-user creation of complex behaviors of AR and IoT}
While being generalizable for creating various stories, one limitation we found about Jigsaw's authoring tool is its expressiveness for defining complex behaviors such as interactivity.
Such complex behaviors typically \lei{require} coding to accomplish.
We encourage future researchers to explore ways to raise the ceiling of expressiveness of the authoring tool for immersive stories while keeping the low barrier of entry for end-users.
For example, in the context of editing animation, visual programming platforms such as Alice~\cite{cooper2000alice} and Scratch~\cite{resnick2009scratch} could be incorporated into authoring tools like Jigsaw to enhance its expressiveness.
In-situ visual programming approaches (e.g., \cite{zhu2023learniotvr, lee2004immersive, flowmatic}) can also be considered as a way to enhance the expressiveness while augmenting the existing intuitive in-situ controls of IoT devices in Jigsaw.
For example, FlowMatic~\cite{zhang2020flowmatic} utilized a visual flow-based diagram that enables end-users to create complex behaviors of virtual objects in VR.
In addition, future researchers can consider ways for reusing or remixing~\cite{dasgupta2016remixing} to make the authoring experience even more expressive for adapting to different storytelling contexts.
The functionality of remixing and reusing could also foster a broader and more vibrant community for exchanging storytelling ideas.

\subsection{Limitations}
Our results demonstrate the benefits and challenges of both the consumption and the creation of immersive stories.
However, our work is not without limitations.
First, Jigsaw currently incorporates a constrained set of modalities including mobile AR, smart lights, fans, and speakers. 
Further research should explore additional modalities like head-worn AR and diverse smart devices to fully uncover the possibilities of immersive storytelling. 
Second, our qualitative evaluation focused on adults; given children represent a key demographic for engaging with and creating stories, future studies could examine their perceptions and involvement. 
\lei{Our particular order of experiencing and re-creating immersive stories in the study could also introduce bias such as learning effects~\cite{fisher1936design}.
For example, users might have learnt what the expected behaviors are prior to the authoring task.
One way to avoid this in the future could be evaluating how end-users can create their own immersive stories, which can offer additional insights on the types of immersive stories that Jigsaw's authoring tool enables as well.}
Follow-up work could also undertake more quantitative investigations into how immersion is impacted by blending virtual and physical augmentations. 
Finally, while designed for storytelling, Jigsaw's adaptability suggests applications in other domains like education, entertainment, and communication. 
Further research might examine how combining IoT and AR could play a role in these and other contexts. 
In summary, our work provides initial insights but warrants expanded investigations into users, technologies, and applications to elucidate the full potential of immersive storytelling.
\section{Conclusion}
In this work, we investigated immersive experiences created by integrating mobile AR with off-the-shelf IoT devices. 
We also introduced an authoring tool for crafting these immersive narratives. 
Our qualitative lab study revealed augmenting IoT devices and enabling user participation can enhance immersion in storytelling, though sensory overload poses a primary challenge. 
We also found our tool's open-ended approach to defining keywords and attributes enabled authoring diverse stories. 
These insights highlight opportunities and obstacles around AR+IoT storytelling. 
Our recommendations for future research and design include: furthering immersion via physical augmentation and participation; enabling immersive story creation by editing adaptable narrative components; and supporting end-user authoring of interactive behaviors. 

\bibliographystyle{ACM-Reference-Format}
\bibliography{jigsaw-chi2023}

\end{document}